\documentclass[12pt]{article}

\usepackage{paper2e}
\usepackage{mydefs2e}

\begin{document}
\renewcommand{\Box}{\Yfund\,}

\newcommand{\N}{$\scr{N}=1$\xspace}

\begin{titlepage}
\preprint{NSF-ITP-99-115; SU-ITP-99/44; UMD-PP-00-031}

\title{Radius Stabilization and\\\medskip
Anomaly-Mediated Supersymmetry Breaking}

\author{Markus A. Luty\footnote{Sloan Fellow.}}

\address{Department of Physics, University of Maryland\\
College Park, Maryland 20742, USA\\
{\tt mluty@physics.umd.edu}}

\author{Raman Sundrum}

\address{Department of Physics, Stanford University\\
Stanford, California 94305-4060, USA\\
{\tt sundrum@leland.stanford.edu}}

\begin{abstract}
We analyze in detail a specific 5-dimensional realization of a
`brane-universe' scenario where the visible and hidden sectors are
localized on spatially separated 3-branes coupled only by
supergravity, with supersymmetry breaking originating in the hidden sector.
Although general power counting allows order $1/M_{\rm Planck}^2$
contact terms between the two sectors in the 4-dimensional theory
from exchange of supergravity Kaluza-Klein modes, we show that
they are not present by carefully matching to the 5-dimensional theory.
We also find that the radius modulus corresponding to the size of the
compactified dimension must be stabilized by additional dynamics in
order to avoid run-away behavior after supersymmetry breaking
and to understand the communication of supersymmetry breaking.
We stabilize the radius by adding two pure Yang--Mills sectors, one
in the bulk and the other localized on a brane.
Gaugino condensation in the 4-dimensional effective theory
generates a superpotential that can naturally fix the radius at
a sufficiently large value that supersymmetry breaking is
communicated dominantly by the recently-discovered mechanism
of anomaly mediation.
The mass of the radius modulus is large compared to $m_{3/2}$.
The stabilization mechanism requires only parameters of order one
at the fundamental scale, with no fine-tuning except for the cosmological
constant.
\end{abstract}

\end{titlepage}

\section{Introduction}
Supersymmetry (SUSY) breaking communicated by supergravity
(SUGRA) is a very natural and attractive solution to
the hierarchy problem.
In its usual incarnation, this mechanism requires only a hidden
sector that breaks SUSY, and the presence in the effective theory
below the Planck scale of the following higher-dimension operators
connecting the hidden and visible sector fields:
\beq\eql{stdhidden}
\bal
\scr{L}_{\rm eff} \sim & \myint d^4\th\, \frac{1}{M_{4}^2}\,
\Si^\dagger \Si
\left[ Q^\dagger Q + (H_u H_d + \hc) \right]
\\
& + \myint d^2\th\, \frac{1}{M_4}
\left[ \Si W^\al W_\al + (\Si^\dagger H_u H_d + \hc)
\right].
\eal\eeq
Here, $M_4$ is the 4-dimensional Planck scale,
$\Si$ is a field in the hidden sector with $\avg{F_\Si} \ne 0$,
$Q$ is a matter field in the visible sector, $H_{u,d}$ are
Higgs fields, and $W_\al$ is a field strength for the standard
model gauge group.
This simple setup generates all required soft SUSY breaking
terms of order $\avg{F_\Si} / M_4 \sim m_{3/2}$ (including the $\mu$
term \cite{GM}).

The main drawback of this scenario is that it does not explain
why the squark masses generated from the term
$\int\! d^4\th\, \Si^\dagger \Si Q^\dagger Q$ approximately
conserve flavor, as required to avoid excessive flavor-changing
neutral currents.
\Ref{RS} proposed an elegant solution to this problem in the context
of higher-dimensional theories.
It was pointed out that if the visible and hidden sectors
are localized on spatially separated `3-branes', then contact terms
of the form \Eq{stdhidden} can be suppressed even though they are
not forbidden by any symmetry of the low-energy theory.
This can be easily understood by focusing on the effective
$D$-dimensional theory ($D  4$) below the string scale $M$, but
above the compactification scale $1/r$.
If the hidden and visible branes are separated
by a distance of order $r$, then the contribution from the
exchange of bulk fields of mass $M \gg 1/r$ is suppressed by the Yukawa
factor $e^{-M r}$.
We also expect the contributions of extended objects with string-scale
tensions to be exponentially suppressed by $e^{-M r}$.
We see that stringy physics generates only exponentially small
contact terms in the $D$-dimensional effective theory.
When we match  the $D$-dimensional theory to the 4-dimensional
low-energy effective theory, a more careful analysis is required to show that
the exchange of supergravity Kaluza-Klein (KK) excitations does not lead
to contact \Kahler terms of order $1/M_4^2$.
We perform this analysis for a specific model in this paper, with the
result that no such terms are generated.
Thus, all of the effective interactions of the form of \Eq{stdhidden} are
highly suppressed, and SUSY breaking must be communicated in a different
way.

\Ref{RS} further argued that, given this suppression of the terms in
\Eq{stdhidden}, the leading contribution to SUSY breaking in the
visible sector arises at loop level, and is directly related to the
conformal anomaly.
This mechanism applied to gaugino masses and $A$ terms was
independently discovered in \Ref{GLMR}, which also gave a detailed
discussion of the exactness of the result.
In this `anomaly-mediated supersymmetry breaking' (AMSB) scenario,
all soft SUSY breaking parameters are completely predicted up to an
overall scale by anomalous dimensions and conserve flavor to a high
degree.
This leads to interesting testable predictions for the gaugino masses
\cite{RS,GLMR,AMphen}.
Unfortunately, the slepton mass-squared terms are predicted to be negative
if the visible sector is the minimal supersymmetric standard
model.
There are have been several suggestions in the literature for natural
solutions to this problem \cite{AMslepton}.

In this paper we investigate the basic features of the AMSB scenario
in detail in a specific 5-dimensional effective field theory.
The theory consists of minimal 5-dimensional SUGRA
compactified on a $S^1 / Z_2$ orbifold.
The two $(3 + 1)$-dimensional boundaries of this space corresponding to the
orbifold fixed-points serve as the
`3-branes' on which the hidden and visible sectors are localized.
The higher SUSY of the 5-dimensional theory is broken explicitly down to \N
in 4 dimensions by the orbifold projection.
This setup is very similar to the five-dimensional effective theory
arising from heterotic M-theory after Calabi-Yau compactification of
six of the eleven dimensions \cite{HW, horava, M, puzzle},
but our field content is the
minimal one required for consistency of the five-dimensional
effective theory.
In particular, the Calabi-Yau moduli do not appear as light fields in our
five-dimensional model. This accounts for the substantial differences
between AMSB and other analyses of supersymmetry breaking in the heterotic
M-theory scenario. We defer consideration of non-minimal field content
for later work.
Our final result is that anomaly mediation is the leading source of
SUSY breaking in the visible sector if the radius is sufficiently
large, but it is crucial to take into account the dynamics of the
radius of the compactified dimension.
While our analysis is limited to a specific 5-dimensional
theory with a particular mechanism for stabilizing the radius,
we believe that these features are more general.

Starting with the 5-dimensional theory described above,
we construct the 4-dimensional effective theory below the
compactification scale to analyze SUSY breaking.
As already mentioned, a crucial feature of the effective theory is
the presence of a radius modulus corresponding to the size of the
compactified dimension.
In particular, if this modulus is not stabilized we will show that
its equations of motion  set to zero the supersymmetry breaking order
parameter for AMSB, namely the four-dimensional SUGRA auxiliary
scalar.
This agrees with a direct five dimensional SUGRA analysis, where
there are no bulk fields which can transmit the effect of such an
order parameter to the visible sector.
This naturally raises doubts as to whether AMSB
occurs in this scenario \cite{puzzle}.
A related issue is the fact that the radius modulus must be
stabilized in order to cancel the cosmological constant in the
presence of SUSY breaking.
We show that if the bare bulk cosmological constant is zero, there is no
potential for the radius modulus, but the low-energy cosmological
constant cannot be cancelled.
In the presence of a bulk cosmological constant,
SUSY breaking gives this modulus a runaway potential.

The picture changes completely
when a stablization mechanism is introduced for the radius.
We propose a stabilization mechanism for the radius modulus that
relies entirely on gaugino condensation and SUSY breaking.
The mechanism requires two super-Yang--Mills (SYM) sectors, one in
the bulk and one localized on a 3-brane, as well as a SUSY
breaking sector localized on the hidden 3-brane.
Upon matching to the 4-dimensional theory, the bulk SYM sector
gives rise to a 4-dimensional SYM sector with a gauge coupling
that depends on the radius $r$.
This gives rise to an $r$-dependent gaugino condensate which,
together with the brane-localized gaugino condensate,
gives a stabilizing potential for the radius modulus.%
\footnote{A similar mechanism can be used to stabilize the radius in
non-supersymmetric theories \cite{nonSUSYstab}.
This may be interesting for solutions of the hierarchy problem
involving extra dimensions that are only slightly larger than the
fundamental scale \cite{RS2}.}
The radius is naturally large compared $M_5$ if the condensation scale of
the 3-brane super-Yang--Mills sector $\La_{\rm bdy}$ is small compared
to $M_5$.
The radius depends only logarithmically on $\La_{\rm bdy}$, we can obtain
a sufficiently large radius for anomaly mediation to dominate if the
theory is strongly coupled near the scale $M_5$.%
\footnote{%
It is interesting that even for strong coupling, this mechanism gives a
radius that is naturally close to the scale where bulk gravitational
loops give a contribution to soft masses comparable to the contributions
of anomaly mediation.
We will not pursue this possibility here.}
Since the condensation scale of the 3-brane super-Yang--Mills
sector is naturally exponentially small compared to the fundamental
scale, this mechanism does not require the introduction of small
parameters at the fundamental scale.
SUSY breaking (and fine-tuning) is required to cancel the net
low-energy cosmological constant.
The mass of the radius modulus is large compared to $m_{3/2}$, and
the effective theory below this scale is of the `sequestered' form
proposed in \Ref{RS}.
The general lesson we draw from this is that
AMSB works provided that moduli are stabilized.
Our stabilization is similar in spirit to the racetrack mechanism \cite{race},
but it does not require large gauge groups and our results follow
from a completely systematic effective field theory analysis.

This paper is organized as follows.
In Section 2 we describe the 5-dimensional model
and carry out the matching to the 4-dimensional effective theory.
We show that there are no $O(1/M_4^2)$ contact terms between the
hidden and visible sectors, and that the cosmological constant
cannot be cancelled in the absence of a mechanism for radius
stabilization.
In Section 3, we show how gaugino condensation can fix the radius,
and show that anomaly mediation works in this scenario.
Section 4 contains our conclusions.

\section{From 5 to 4 Dimensions}
\subsection{The 5-dimensional Model}
We consider minimal (ungauged) 5-dimensional SUGRA
compactified on a $S^1 / Z_2$ orbifold with matter and gauge fields
localized on the two orbifold boundaries.
This system is relatively simple to study because the orbifold
projection explicitly breaks the supersymmetry of the 5-dimensional
theory (8 real supercharges)  down to \N in 4 dimensions
(4 real supercharges).

The on-shell lagrangian for the bosonic fields of 5-dimensional
SUGRA is \cite{sugra5}
\beq\eql{SUGRA5}
\bal
\scr{L}_{\rm SUGRA,5} = -M_5^3 & \biggl[ \sqrt{-g^{(5)}} \left(
\sfrac{1}{2} \scr{R}^{(5)}
+ \sfrac{1}{4} H^{MN} H_{MN} \right)
\\
& \quad
+ \frac{1}{6\sqrt{6}} \ep^{MNPQR} B_M H_{NP} H_{QR}
+ \hbox{\rm fermion\ terms} \biggr],
\eal\eeq
where $M, N, \ldots = 0, \ldots, 3, 5,$ are 5-dimensional
spacetime indices, and $H_{MN} = \partial_M B_N - \partial_M B_N$
is the field strength for the graviphoton $B_M$.
Under the $Z_2$ parity, the fields transform as
$\phi(x^5) \mapsto \pm \phi(-x^5)$, where the parity assignments
of the bosonic fields are given in Table~1.
The orbifold projection keeps only those field configurations
that are even under $Z_2$.

We assume that there are fields localized on the orbifold boundaries,
so these must be coupled to SUGRA.
The lagrangian has the form
\beq
\scr{L}_5 = \scr{L}_{\rm SUGRA,5}
+ \de(x^5) \scr{L}_{\rm vis} + \de(x^5 - \pi r) \scr{L}_{\rm hid}.
\eeq
We will not need the details of the bulk-boundary couplings in
$\scr{L}_{\rm vis}$ and $\scr{L}_{\rm hid}$, but it is important for
us to know that such couplings exist and preserve \N SUSY.
As shown in \Ref{MP} for 5-dimensional gauge- and hypermultiplets,
the couplings of bulk and boundary fields can be worked out in a
straightforward fashion if the auxiliary fields of the bulk theory
are known.
Building on earlier work \cite{early5}, an explicit off-shell formulation
for 5-dimensional SUGRA was recently given by Zucker \cite{zucker}.
Following \Ref{MP}, one first decomposes the 5-dimensional SUGRA
multiplet into off-shell multiplets of the unbroken 4-dimensional \N
SUSY.
In addition to the \N SUGRA multiplet, this yields two vector
multiplets (with vector fields $g_{5\mu}$ and $B_\mu$) with odd
orbifold parity, and one chiral multiplet (with real scalar fields
$g_{55}$ and $B_5$) with even parity.
It should then be possible to couple these multiplets to \N fields
localized on the boundaries using the usual \N superfield calculus.

\begin{table}[t] 
\centering
\begin{tabular}{c|c}
Field & $Z_2$ Parity \\
\hline
$g_{\mu\nu}$ & $+$ \\
$g_{5\mu}$ & $-$ \\
$g_{55}$ & $+$ \\
$B_\mu$ & $-$ \\
$B_5$ & $+$ \\
\end{tabular}
\\
\capt{Bosonic fields of 5-dimensional SUGRA with
their $Z_2$ parity assignments.
The parity assignments of the graviphoton fields are fixed
by the the Chern--Simons term.}
\end{table}

\subsection{Matching to 4 Dimensions}
We now consider integrating out the KK modes of the
5-dimensional SUGRA multiplet at the scale $r$ to obtain
a 4-dimensional effective theory.
We are interested in effects of order $1/M_4^2 \sim 1  / (r M_5^3)$,
which means that we can restrict attention to tree-level effects
in the SUGRA fields.
(In the normalization of the SUGRA fields given in \Eq{SUGRA5},
the propagator for all bosonic SUGRA fields is of order
$1/M_5^3$.)
There are SUGRA loop effects suppressed by additional powers of
$1/(M_5 r)^3 \sim 1/(M_4 r)^2$. For some values of $r$ these effects could
be interesting \cite{RS}. Here we will simply
assume that $r$ is sufficiently large that these loop effects can be
neglected.
The matching of the SUGRA fields at tree-level is performed simply
by using the metric
\beq
ds^2 = g_{\mu\nu}(x) dx^\mu dx^\nu + r^2(x) d\vartheta^2,
\eeq
where $\vartheta \in [0, \pi]$ is a coordinate for the compact dimension,
and $g_{\mu\nu}(x)$, $r(x)$ parameterize the massless metric and
radius modulus fields.
(We are implicitly expanding about a flat metric, so the zero-mass
KK modes are independent of $\vartheta$.)
Ignoring the boundary fields for the moment, the bosonic terms in the
4-dimensional effective theory are
\beq\eql{eff4}
\scr{L}_4 = -2\pi M_5^3 \sqrt{-g^{(4)}} \left[
\frac{r}{2} \scr{R}^{(4)}
+ \frac{1}{2 r} \partial^\mu B_\vartheta \partial_\mu B_\vartheta \right].
\eeq
Note that there is no explicit kinetic term for the radius modulus.
After an $r$-dependent Weyl rescaling of the metric, a kinetic term
for the radius modulus is generated (with the correct sign).
The couplings of the radius modulus to boundary fields is very
different in the two bases.
Before Weyl rescaling, there are no couplings of $r$ to boundary
fields at leading order in the low-energy expansion.
This is because $r$ arises from fluctuations of $g_{55}$,
which by general covariance can only couple to the $55$ component of
the matter stress tensor. This component vanishes for matter confined to
3-branes, at leading order in $1/M_5$.
At higher order in derivatives and $1/M_5$, we can write terms containing the
curvature tensor that depend on derivatives of $r$ but these will be a small
correction.
In the rescaled basis the radius modulus has
non-derivative couplings to fields localized on the branes.

\Eq{eff4} is to be matched to the most general lagrangian
describing 4-dimensional SUGRA coupled to a modulus $T$.
Using the superconformal approach to SUGRA \cite{conformalSUGRA},
this can be written as
\beq\eql{SUGRA4}
\scr{L}_{\rm SUGRA,4} = \myint d^4\th\, \phi^\dagger \phi\, f(T^\dagger, T).
\eeq
where
\beq
\phi = 1 + \th^2 F_\phi
\eeq
is the conformal compensator.
We do not include a superpotential in \Eq{SUGRA4} because $T$ has no
potential in this approximation.
(Recall that we are not including a bulk cosmological constant.)
After integrating out the auxiliary fields, the bosonic terms of \Eq{SUGRA4}
are
\beq[genSUGRA4]
\bal
\scr{L}_{\rm SUGRA,4} = \sqrt{-g^{(4)}} \biggl[ &
\frac{1}{6} f \scr{R}^{(4)}
- \frac{1}{4 f} (f_T \partial^\mu T - \hc) (f_T \partial_\mu T - \hc)
\\
&
- f_{T^\dagger T} \partial^\mu T^\dagger \partial_\mu T
+ \hbox{\rm fermion\ terms} \biggr],
\eal\eeq
where $f_T = \partial f / \partial T$, {\it etc\/}.
An important point is that \Eq{genSUGRA4} must be matched to \Eq{eff4}
\emph{without} Weyl rescaling.
The reason is that if boundary fields are included, the theory expressed
in terms of the Weyl-rescaled metric contains only non-derivative couplings
to the radius modulus.
\Kahler terms involving both $T$ and boundary terms necessarily contain
derivative interactions of $T$, the only consistent way to match is if
the \Kahler terms are $T$-independent.
\Eq{eff4} then shows that there is an explicit kinetic term for only one of
the real scalar fields in $T$, so we must have $f_{T^\dagger T} \equiv 0$.
This implies that $f$ is the sum of a holomorphic plus
antiholomorphic function, so we can make a field redefinition so that
$f = -M_5^3 \cdot (T + T^\dagger)$.
Writing $T = T_1 + i T_2$, we have
\beq\eql{noscale}
\scr{L}_{\rm SUGRA,4} &= -M_5^3 \myint d^4\th\,
\phi^\dagger \phi\, (T + T^\dagger)
\\
\eql{general}
&= -M_5^3 \sqrt{-g^{(4)}} \left[
\frac{T_1}{3} \scr{R}^{(4)}
+ \frac{1}{2 T_1} \partial^\mu T_2 \partial_\mu T_2 +
\hbox{fermion\ terms} \right].
\eeq
Comparing this 
with \Eq{eff4}, we can identify
\beq\eql{Tid}
\Re(T) = 3 \pi r,
\qquad
\Im(T) = \sqrt{6} \pi B_\vartheta.
\eeq
\Eq{noscale} has  the `no-scale' form considered long ago
\cite{noscale}.
The essential new ingredient in the present case is that the no-scale
form is stable under radiative corrections because the cutoff of the
4-dimensional theory is of order $1/r \ll M_4$.


We now consider the fields localized on the orbifold boundaries.
We are particularly interested in contact interactions between
the hidden and visible sectors.
The only contact interaction of order $1/M_4^2$ in the 4-dimensional
effective theory that is not forbidden by symmetries is
\beq\eql{badop}
\frac{1}{M_4^2} \myint d^4\th\, (\Si^\dagger \Si) (Q^\dagger Q)
= \frac{4}{M_4^2} (\psi_\Si \psi_Q) (\bar{\psi}_\Si \bar{\psi}_Q)
+ \cdots
\eeq
where we have explicitly shown the 4-fermion component.
The only diagrams that can contribute to the 4-fermion term in
\Eq{badop} at order $1/M_4^2$ consist of tree-level exchange of
bosonic SUGRA fields.
The bulk-boundary couplings cannot involve any suppression
by $1/M_5$, otherwise the final result will be less than $1/M_4^2$.
It may appear that these conclusions are invalidated by
power-divergent loop graphs with a cutoff of order $M_5$.
However, general renormalization theory tells us that the divergent
contributions will have the same structure as local terms in the
effective field theory, and therefore do not give new effects.

Now, the exchange of Kaluza-Klein excitations of the graviton couple to
derivatives of the fermion fields and therefore cannot yield a term of
the form \Eq{badop}.
Couplings of the graviphoton to boundary fields are restricted by the
orbifold projection and graviphoton gauge invariance
\beq
\de B_M = \partial_M \al,
\qquad
\al(-x^5) = -\al(x^5).
\eeq
Boundary fields cannot be charged under this symmetry because $B_\mu$
vanishes on the boundary.
The only term consistent with these constraints that can give rise
to the 4-fermion term in \Eq{badop} has the form
\beq\eql{bdyterm}
\De\scr{L}_5 =
\de(x^5) H_{5\mu} K_{\rm vis}^\mu
+ \de(x^5 - \pi r) H_{5\mu} K_{\rm hid}^\mu
\eeq
where $K^\mu$ is a dimension-3 current constructed from boundary fields;
its precise form will be determined by matching to the 4-dimensional theory.

The power-counting argument above shows that \Eq{bdyterm} will give
rise to contact terms of order $1/M_4^2$ from tree-level exchange of
$B_\mu$ fields.
We can determine these terms by integrating out $B_\mu$ using its
classical equations of motion.
Imposing periodicity and consistency with the orbifold projection,
we obtain
\beq
\partial_5 B^\mu = \frac{1}{M_5^3} \left[
\de(y) K_{\rm vis}^\mu + \de(y - \pi r) K_{\rm hid}^\mu
- \frac{1}{2\pi r} (K_{\rm vis} + K_{\rm hid})^\mu \right].
\eeq
In this computation it was important that we considered the
$B_{\vartheta}$ field to be independent of $x_5$,
   corresponding to the zero-mode
(Im$T$) of the five-dimensional field.
Substituting back into the lagrangian and integrating over the compact
dimension to obtain the 4-dimensional effective theory,
we obtain the contact terms%
\footnote{This procedure also gives rise to terms proportional to
$\de(0) \cdot (K_{\rm vis}^2 + K_{\rm hid}^2)$ in the 4-dimensional
effective theory;
these are cancelled by boundary terms proportional to $\de(0)$ in the
5-dimensional theory.
For a discussion of the origin of these terms, see \Refs{horava,MP}.}
\beq[contact5]
\De\scr{L}_4 = -\frac{1}{r} \partial_\mu B_{\vartheta}
(K_{\rm vis} + K_{\rm hid})^\mu
- \frac{1}{4 \pi M_5^3 r} (K_{\rm vis} + K_{\rm hid})^\mu
(K_{\rm vis} + K_{\rm hid})_\mu.
\eeq
We compare this with the contact terms in the 4-dimensional SUGRA
with matter fields:
\beq
\scr{L}_4 = \myint d^4\th\, \phi^\dagger \phi \left[
-M_5^3 (T + T^\dagger) + f_{\rm vis} + f_{\rm hid} \right].
\eeq
As argued above, $f_{\rm vis}$ and $f_{\rm hid}$ are independent of
$T$ because any dependence would imply a coupling of $r$ (and hence $g_{55}$)
to brane
fields (without Weyl rescaling).
We therefore obtain
\beq
\eql{contact4}
\scr{L}_4 = -\frac{1}{2 T_1} \partial_\mu T_2 J^\mu
- \frac{1}{8 M_5^3 T_1} J^\mu J_\mu + \cdots.
\eeq
Here $J^\mu = J_{\rm vis}^\mu + J_{\rm hid}^\mu$ with
\beq
J^\mu = i (f_a \partial^\mu \phi^a - \hc)
+ f^a{}_b \psi_a \si^\mu \bar{\psi}^b,
\eeq
where
$f^a{}_b = \partial^2 f / (\partial \Phi^\dagger_a \partial\Phi^b)$,
{\it etc\/}.

Comparing \Eqs{contact5} and \eq{contact4} and using
\Eq{Tid}, we see that matching the $\partial_\mu B_{\vartheta} K^\mu$
term requires
\beq
K^\mu = \frac{1}{\sqrt{6}} J^\mu.
\eeq
With this identification, the $J^\mu J_\mu$ contact terms also match.
This matching would be spoiled by additional contact terms of the form
\Eq{badop}, so we conclude that these operators are absent in the
4-dimensional effective theory.

Putting together the various pieces, the four-dimensional effective theory
below the compactification scale has the general form,
\beq[sequest]
\scr{L}_4 = -M_5^3 \myint d^4\th\, \phi^\dagger \phi\, (T + T^\dagger)
+ \scr{L}_{\rm hid} + \scr{L}_{\rm vis},
\eeq
where $\scr{L}_{\rm hid}$ is made out of only hidden sector and
four-dimensional supergravity (off-shell) multiplets and
$\scr{L}_{\rm vis}$ is made out of only visible sector and
four-dimensional supergravity multiplets. Both are independent of the $T$
chiral multiplet.

\subsection{The Role of the Radius}
We now consider SUSY breaking on the hidden-sector boundary
in the theory above.
We will show that the presence of an unstabilized radius modulus
gives rise to severe difficulties in this scenario when SUSY is broken
in the hidden sector.

Independently of how SUSY is broken, it is easy to see from
\Eq{sequest} that the $F_T$ equation of motion sets $F_{\phi} = 0$.
This implies that there are no contact terms between the visible and
hidden sectors in this theory, consistent with the fact that there is
no propagating bulk scalar field in the SUGRA multiplet that could mediate
such
terms.
This makes it rather mysterious how SUSY breaking can be communicated
from the hidden to the visible sector \cite{puzzle}, especially since
$F_\phi$ is the order parameter for AMSB in the visible sector
\cite{RS,GLMR}.

This feature also gives rise to difficulties in cancelling the cosmological
constant.
SUSY breaking on the hidden-sector boundary gives rise to a nonzero
vacuum energy independent of the radius modulus $T$.
In generic four-dimensional SUGRA models this positive contribution
to the cosmological constant can be cancelled by negative SUGRA
contributions arising from $F_{\phi} \neq 0$,
but this mechanism is clearly not available here.
One can attempt to remedy this by adding a SUSY-preserving
five-dimensional cosmological constant to the theory.
To linear order in the cosmological constant, the effect of this is
to add a superpotential term linear in $T$ to \Eq{sequest}.
The potential arising from this theory is now
\beq\eql{badradeffpot}
V = -\frac{k}{M_5^3} \Re(T) + V_{\rm hid},
\eeq
where $V_{\rm hid}  0$ is the vacuum energy from hidden sector SUSY
breaking, and $k  0$ sets the size of the bulk cosmological constant.
However, this introduces a new problem, namely runaway behavior for the
radius modulus.%
\footnote{When $T$ becomes sufficiently large, the linearized
approximation for the effect of a bulk cosmological constant is no
longer valid.
We have checked that including the full non-linear effects does not
stop the runaway behavior.}
We see that we cannot obtain an appropriate setting for
AMSB without adding new physics to stabilize the modulus.

We mention that another means of breaking SUSY is to {\it not} have a
hidden sector which breaks SUSY by itself but rather to simply have a constant
superpotential on a brane (and no bulk cosmological constant).
Then one finds that $F_T \neq 0$, but $F_{\phi} = 0$, so SUSY is
broken but the cosmological constant vanishes.
This is the basic no-scale mechanism of SUSY breaking \cite{noscale}.
We do not pursue this scenario here because it involves the vanishing
of the AMSB order parameter $F_{\phi}$.

\section{Radius Stabilization}
We now show that the problems found above are solved by
dynamically stabilizing the modulus.
This modulus must be stabilized in any case for phenomenological
reasons.
(The radius modulus must have a mass larger than of order
$1\cm^{-1}$ to avoid conflict with post-Newtonian tests of
gravity \cite{PN}.)
We will focus on a specific mechanism for stabilizing the radius
modulus that requires only a super-Yang--Mills (SYM) sector in the bulk,
and another SYM sector on one of the boundaries.
We assume that the bulk cosmological constant is negligible;
this is natural because of the presence of bulk SUSY.

\subsection{Bulk Super-Yang--Mills}
We begin by discussing the bulk SYM sector.
At the compactification scale $1/r$, this theory matches onto a
4-dimensional SYM theory with a gauge coupling that depends on
$r$.
The scale where the effective 4-dimensional SYM theory becomes strong
therefore depends on $r$, and gaugino condensation generates a
dynamical superpotential that depends on the modulus $T$.
%
%
The fact that the dynamical superpotential for $T$ is generated by
supersymmetric dynamics rather than induced by SUSY breaking in the
hidden sector allows the mass of the modulus to be large compared
to $m_{3/2}$.
This means that below the scale of the radius modulus, the effective
theory has the `sequestered' form discussed in \Ref{RS}, and
the leading contribution to SUSY breaking in the visible sector
comes from anomaly mediation.

\begin{table}[t] 
\centering
\begin{tabular}{c|c}
Field & $Z_2$ Parity \\
\hline
$A_\mu$ & $+$ \\
$A_5$ & $-$ \\
$\Phi$ & $-$ \\
$\la^1$ & $+$ \\
$\la^2$ & $-$ \\
\end{tabular}
\\
\capt{Fields of 5-dimensional super-Yang--Mills sector with
their $Z_2$ parity assignments.}
\end{table}

The bulk SYM multiplet consists of a vector field $A_M$, a real scalar
$\Phi$, and a symplectic Majorana gaugino $\la^j$ ($j = 1, 2$).
These fields are taken to transform under the orbifold projection
as shown in Table 2.
The even fields form an \N SYM multiplet $\scr{V}$, while the odd
fields form an \N chiral multiplet $\Psi$.
These fields can be coupled to the boundary fields using the usual rules
for constructing \N invariants.
(For more details, see \Ref{MP}.)

We assume that the fields on the boundaries are uncharged under
the bulk SYM sector.
However, there are in general higher-dimension operators
coupling the bulk SYM fields to the boundary fields.
Using a normalization of the fields where the gauge coupling
is factored out of the kinetic terms
\beq
\scr{L}_5 = \frac{1}{g_5^2} \tr \left[
-\frac{1}{4} F^{MN} F_{MN} + \partial^M \Phi \partial_M \Phi
+ \cdots \right],
\eeq
the bulk SYM propagator is proportional to $g_5^2 \sim 1/M_5$.
Therefore, exchange of SYM fields between the boundaries can
give rise to contact terms of order
$1/M_4^2 \sim 1/(r M_5^3)$ only if there are boundary couplings
of order $1/M_5$.
However, it is easy to see that no such terms are possible unless
there is a singlet $S$ on the boundary, in which case we can write
\beq
\De\scr{L}_5 = \de(y) \myint d^2\th\,
\frac{1}{M_5} S \tr(\scr{W}^\al \scr{W}_\al)
+ \hc,
\eeq
where $\scr{W}^\al$ is the field strength of the \N SYM field $\scr{V}$.
(Note that boundary couplings involving the \N chiral multiplet $\Psi$ are
restricted by gauge invariance
$\de\Psi = i\partial_5 \al$,
$\al(-x^5) = +\al(x^5)$.)
If there are singlets in both the hidden and visible sector, this
will induce contact terms between them only at the 1-loop level, and
the presence of two SYM propagators in the leading diagram
means that the effects are suppressed
by $1/M_5^4$, and therefore negligible.
(The contact terms are \Kahler terms by $U(1)_R$ invariance.)
We conclude that introducing the bulk SYM sector does not introduce
new contact terms into the effective 4-dimensional theory.

We now construct the 4-dimensional effective theory for the bulk SYM sector.
When we perform the KK decomposition, the odd fields have
KK masses starting at $1/r$, and are therefore integrated
out.
The even fields have a massless zero mode, which becomes a
4-dimensional SYM sector in the effective theory.
The tree-level matching condition for the effective 4-dimensional
gauge coupling is
\beq
\frac{1}{g_4^2} = \frac{2 \pi r}{g_5^2}.
\eeq
Because $g_4$ depends on $r$, gaugino condensation in the effective
4-dimensional SYM sector will give rise to a $T$-dependent dynamical
superpotential.

The $T$ dependence of the dynamical superpotential can be
determined exactly using holomorphy arguments \cite{KL}.
The holomorphic 4-dimensional gauge coupling
$S = 1/(2 g_4^2) + \cdots$
is given exactly by
\beq\eql{exactmatch}
S(\mu = 1/g_5^2) = \frac{2 T}{3 g_5^2} + c,
\eeq
where $c$ is a real constant that parameterizes the scheme dependence.
It may appear that cancelling large logs requires us to match
at a scale $\mu \sim 1/r$:
\beq
S(\mu = 1/T) \stackrel{\textstyle ?}{=} \frac{2T}{3 g_5^2} + c.
\eeq
However, for $\mu < 1/r$ this leads to (for an $SU(N)$ gauge group)
\beq
S(\mu) \stackrel{\textstyle ?}{=} \frac{2 T}{3 g_5^2}
+ \frac{3N}{16\pi^2} \ln(\mu T) + c.
\eeq
The logarithmic dependence on $T$ implies that
$1/g_4^2 \propto \Re(S)$ depends on $\Im(T) \propto B_\vartheta$.
But from the 5-dimensional theory we know that $B_\vartheta$ is
derivatively coupled, so this is impossible.
It is easy to see that
the only way to avoid this contradiction consistent with holomorphy is
\Eq{exactmatch}.
We have also checked that carefully evaluating the threshold
corrections due to the infinite tower of SYM KK states also reproduces
\Eq{exactmatch}.
The dynamical scale of the theory is therefore
\beq\eql{Laprop}
\La_{\rm bulk} \propto \frac{1}{g_5^2} e^{-32\pi^2 T / (9 N g_5^2)}.
\eeq

In order to obtain believable numerical estimates
we need to estimate the constant of proportionality in \Eq{Laprop}.
This can be done using `\naive dimensional analysis' (NDA)
\cite{NDA,SUSYNDA}.%
\footnote{NDA is applied to higher-dimension theories with branes
in \Ref{U1}.}
The principle of NDA is that in a strongly-coupled theory with
no small parameters, both the fundamental and the effective
theory become strongly coupled (in the sense that loop corrections
are order 1) at the same scale.
To estimate $\La$, note that NDA implies that if the gauge
coupling and the radius are chosen so that the fundamental theory
is strongly-coupled at a scale $\La_0$, then $\La \sim \La_0$.
The strong-coupling value of the 5-dimensional gauge coupling is
\beq\eql{gstrong}
\left. g_5^2 \right|_{\rm strong} \sim \frac{\ell_5}{N \La_0},
\eeq
where $\ell_5 = 24\pi^3$ is the (inverse of the) 5-dimensional loop counting
parameter, and we have taken into account the $N$ dependence
appropriate for the large $N$ limit.
The strong-coupling value of the radius is where the KK modes
have mass of order $\La_0$:
\beq
\left. r \right|_{\rm strong} =
\frac{1}{3\pi} \left. T \right|_{\rm strong}
\sim \frac{1}{\La_0}.
\eeq
This implies that the strong-coupling value of the exponential in
\Eq{Laprop} is order 1, and we obtain
\beq
\La_{\rm bulk} \sim \frac{\ell_5}{N g_5^2}
e^{-32\pi^2 T / (9 N g_5^2)}.
\eeq
The dynamical superpotential generated in the
4-dimensional effective theory is therefore 
%
\beq
W_{\rm bulk,dyn} \sim \frac{1}{N \ell_4} \La_{\rm bulk}^3
\sim \frac{\ell_5^3}{\ell_4 N^4 g_5^6} e^{-32 \pi^3 T / (3 N g_5^2)}.
\eeq
Using $\ell_5 = 24\pi^3$ and $\ell_4 = 16\pi^2$, the
dimensionless prefactor is
$\ell_5^3 / \ell_4 = 864 \pi^7 \simeq 3 \times 10^{6}$.
However, this estimate depends sensitively on the value
used for $\ell_5$, and should be regarded as very uncertain.
Nonetheless, it is clear that the prefactor will be large unless
NDA is completely misleading.%
\footnote{In this connection, it may be worthwhile to point out
that exact results obtained in $\scr{N} = 2$ theories spectacularly
confirm the expectations of NDA \cite{N2NDA}.}

\subsection{Boundary Super-Yang--Mills}
In addition to the bulk SYM sector, we assume that the theory contains a SYM
sector localized on one of the boundaries.
As with the bulk SYM, we assume that there are no matter fields
charged under the SYM gauge group.
If this SYM sector is in the hidden sector, there is no danger
from flavor-violating higher-dimension contact terms.
If it is in the visible sector, the lowest-dimension potentially
flavor-violating operator is
is
\beq
\De\scr{L}_5 \sim \de(y) \myint d^4\th\, \frac{1}{M_5^3}
Q^\dagger Q \tr(W^\al W_\al) + \hc,
\eeq
where $W^\al$ is the field strength of the boundary SYM multiplet.
This gives flavor-violating interactions suppressed by
$(\La_{\rm bdy} / M_5)^3$, where $\La_{\rm bdy}$ is the dynamical
scale.
This is negligible for the values of $\La_{\rm bdy}$ we will be
interested in (see below), and we conclude that the boundary SYM
sector may be either in the hidden or the visible sector.

\subsection{4-Dimensional Effective Theory}
Now we are ready to analyze the 4-dimensional effective theory,
including all sectors.
Below the scale $1/r$, the 4-dimensional theory consists of 4-dimensional
SUGRA and the modulus $T$ coupled to the bulk and boundary SYM sectors.
In addition, the theory contains the visible and hidden
sectors, which we do not specify explicitly.
We now write the effective lagrangian
below the scales $\La_{\rm bulk}$ and $\La_{\rm bdy}$ where
the SYM sectors become strong, and below the scale of SUSY
breaking in the hidden sector.
In this regime, the only light fields are the SUGRA fields,
the modulus $T$, the Goldstino from the SUSY breaking sector,
and the visible sector fields.
The effective lagrangian is
\beq\eql{theLeff}
\bal
\scr{L}_{\rm eff} = & -M_5^3 \myint d^4\th\,
\phi^\dagger \phi\, (T + T^\dagger)
\\
& + \left( \myint d^2\th\, \phi^3
\left[ c + a e^{-b T} \right] + \hc \right)
- V_{\rm hid} + \cdots.
\eal\eeq
Here
\beq
c \sim \frac{1}{\ell_4} \La_{\rm bdy}^3
\eeq
arises from gaugino condensation in the boundary SYM theory
(we neglect $N$ dependence in the boundary SYM theory);
\beq\eql{abdef}
a \sim \frac{\ell_5^3}{\ell_4 N^4 g_5^6},
\qquad
b = \frac{32\pi^2}{3 N g_5^2}
\eeq
arise from gaugino condensation in the bulk SYM theory;
and $V_{\rm hid} > 0$ is the vacuum energy generated by the SUSY breaking
sector.
We have chosen not to add a 5-dimensional cosmological constant.
The constant $c$ can be chosen real by a $U(1)_R$ rotation, but
$a$ is in general complex.
The terms omitted in \Eq{theLeff} contain the interactions of the
visible sector fields and a Goldstino from SUSY breaking in the
hidden sector (which will eventually become the longitudinal
components of the massive gravitino).
The terms involving the Goldstino can be included using a non-linear
realization of SUSY coupled to SUGRA \cite{nonlinSUSY}, but are not
relevant for computing the effective potential for $T$;
the same is true for the visible sector interactions.

The superpotential in \eq{theLeff} is exact, but the \Kahler
potential contains unknown $\scr{O}(1/M_4^4)$ corrections from loop
corrections and higher-dimension operators.
These will be shown to give small corrections below.

We now turn to the minimization of the scalar potential, neglecting
corrections to the \Kahler potential.
The scalar potential obtained from \Eq{theLeff} is
\beq\eql{effpot}
\!\!\!\!\!\!
V = \frac{1}{M_5^3} \left\{
\left( 3 c^* b a e^{-b T} + \hc \right)
+ b \left[ b (T + T^\dagger) + 6 \right] |a|^2 e^{-b (T + T^\dagger)}
\right\}
+ V_{\rm hid}.
\eeq
Note that the first term is proportional to the boundary SYM gaugino
condensate.
Only the first term in \Eq{effpot} depends on $\Im(T)$.
Minimizing with respect to $\Im(T)$, we obtain the
effective potential for $T_1 = \Re(T)$:
\beq
V = \frac{1}{M_5^3} \left\{ -6 b |a| |c| e^{-b T_1}
+ 2 b (b T_1 + 3) |a|^2 e^{-2 b T_1} \right\}
+ V_{\rm hid}.
\eeq
The term in brackets is a sum of two different exponentials with
opposite signs, the negative sign in the first term arising from
the minimization with respect to $\Im(T)$.
As $T_1 \to \infty$ the first term dominates, and the potential
approaches $+V_{\rm hid}$ from below.
Provided the second term dominates for small $T_1$ there will
be a nontrivial minimum with vacuum energy \emph{below} $+V_{\rm hid}$.
This means that the parameters can be adjusted to give a vanishing
cosmological constant.

We look for a minimum with $b \avg{T_1} \gg 1$.
Explicitly carrying out the minimization we find that
\beq\eql{r}
b \avg{T_1} e^{-b \avg{T_1}} \simeq \frac{3 |c|}{2 |a|}
\sim \frac{N^4 \La_{\rm bdy}^3 g_5^6}{\ell_4 \ell_5^3},
\eeq
where we have neglected terms suppressed by powers of $1/(b \avg{T_1})$.
Note that $\avg{T_1}$ can be made arbitrarily large
by making $\La_{\rm bdy}$ small compared to $1/g_5^2$.
(The loop suppression factors also tend to increase $\avg{T_1}$.)
The vacuum energy at the minimum is
\beq
\avg{V} \simeq -\frac{3 |c|^2}{M_4^2} + V_{\rm hid},
\eeq
where $M_4^2 = M_5^3 \pi r$.
The fact that the first term is negative allows us to choose the
parameters to fine-tune the cosmological constant to zero.

Because the superpotential has non-trivial $T$ dependence, the $F_T$
equation of motion no longer sets $F_\phi = 0$.
Instead we have
\beq\eql{F}
\avg{F_\phi} \simeq \frac{|c|}{M_4^2}.
\eeq
SUSY is broken, and the gravitino mass is
\beq\eql{gravitino}
m_{3/2} \sim \frac{V_{\rm hid}^{1/2}}{M_4} \sim \frac{|c|}{M_4^2},
\eeq
so that $\avg{F_\phi} \sim m_{3/2}$.

The mass of the radius modulus is computed from
\beq
\avg{V''} \simeq \frac{6 b^2 |c|^2}{M_4^2},
\eeq
where the primes denote differentiation with respect to $T_1$.
The kinetic term for $T_1$ arises from mixing with the metric;
it can be made manifest by making a $T_1$-dependent Weyl transformation.
This gives a kinetic term $\sim M_4^2 (\partial T_1)^2 / T_1^2$,
and the physical mass of the radius modulus is
\beq
m_{r}^2 \sim \frac{b^2 |c|^2}{M_4^4} \avg{r}^2.
\eeq
It is easy to see that the other real scalar and the fermion component of
$T$ also get a mass of this order.
Comparing with \Eq{gravitino}, we see that
\beq
\frac{m_{r}}{m_{3/2}} \sim b \avg{r} \gg 1.
\eeq
Since the modulus is heavy we can integrate it out of our effective theory.
The different component fields in $T$ have mass differences of order
$m_r$, so this is not an approximately supersymmetric threshold;
also it is easy to see that $F_T$ does not vanish
($\avg{F_T} \sim \avg{r} \avg{F_{\phi}}$).
However, $T$ couples to visible sector only through higher-dimension
derivative interactions (recall that the modulus is the zero-mode of the
five-dimensional graviton polarized transverse to the branes), so
this does not give a contribution to SUSY breaking in the visible
sector at order $1/M_4^2$.
We conclude that at order $1/M_4^2$, the effective theory below the
modulus mass is precisely the `sequestered form' proposed in
\Ref{RS}: the visible sector is coupled only to a geometrically flat
four-dimensional SUGRA background with broken SUSY
($F_{\phi} \neq 0$).

We now return to the question of the corrections to the \Kahler
potential in \Eq{theLeff}.
The \Kahler potential contains unknown $\scr{O}(1/M_4^4)$
corrections from loop corrections and higher-dimension operators,
and one might worry  that these are more important than the exponentially
(in $T$) suppressed effects in the superpotential.
This does not occur because the potential vanishes in the limit
where the superpotential vanishes, so the \Kahler corrections
enter multiplicatively.
This  ensures the stability of the results above, in that the \Kahler
corrections to the modulus potential are of order $1/(r M_5)$ smaller
than the leading potential we computed.

We now show that this scenario for radius stabilization can give rise to
a sufficiently large radius without introducing small numbers or fine
tuning.
 From \Eq{r}, the stabilized value of the radius is
\beq\eql{logr}
r \sim \frac{N g_5^2}{\ell_5} \ln\left(
\frac{\ell_5^3}{M_5^2 \avg{F_\phi} N (N g_5^2)^3} \right).
\eeq
Because the radius depends logarithmically on the
fundamental parameters, we cannot obtain hierarchies of many
orders of magnitude.
In fact, because of the factor $1/\ell_5 \sim 10^{-3}$ multiplying
the logarithm in \Eq{logr}, the bulk SYM gauge coupling $g_5$ must
be large, and the fundamental theory must be close to strongly coupling.

The simplest assumption is that both gravity and the bulk SYM sector
become strong at a single scale $\La_0$.
NDA gives $\La_0 \sim (\ell_5)^{1/3} M_5 \sim 10 M_5$, and we will
take this scale to be the fundamental scale of the theory (\eg the
scale of string/M-theory excited states).
Using the NDA estimates for $\La_0$ and $g_5^2$, we obtain
\beq
r \sim \frac{1}{\La_0} \ln \left( \frac{\La_0}{N \avg{F_\phi}} \right).
\eeq
Using $\avg{F_\phi} \sim 100\TeV$ and
$\ell_4 M_4^2 \sim \ell_5 M_5^3 r$, we obtain (for $N = 2$)
\beq
r \sim \frac{30}{\La_0},
\qquad
\La_0 \sim 2 \times 10^{18}\GeV.
\eeq
This is sufficient to suppress FCNC effects from massive string
states, but bulk gravitational loops give contact terms suppressed
by \cite{RS}
\beq
\frac{1}{\ell_4 M_4^2 r^2} \sim \frac{1}{\ell_5 M_5^3 r^3}
\sim \frac{1}{\La_0^3 r^3} \sim 4 \times 10^{-5}.
\eeq
This gives a contribution to soft scalar mass-squared terms of order
$\avg{F_\phi}^2 / (\La_0 r)^3 \sim (600\GeV)^2$,
which is comparable to the contribution from anomaly mediation!
It is interesting that this mechanism for radius stabilization
can naturally stabilize the radius at a value where loop effects
are important.
This may give a solution to the problem of negative slepton masses,
but we will not pursue this point here.

Another possibility is that the bulk SYM sector becomes strong at a
scale $\La_{\rm gauge}$ that is smaller than the scale
$\La_{\rm grav}$.
Here, $\La_{\rm gauge}$ is a fundamental scale of new strong physics, while 
$\La_{\rm grav}$  is not directly a physical scale, but corresponds to
a weak gravitational coupling at the fundamental scale $\La_{\rm gauge}$.
This occurs naturally if the gauge interactions propagate in
fewer dimensions than gravity in the fundamental theory.
For $\La_{\rm gauge} / \La_{\rm grav} \sim \sfrac{1}{10}$,
we obtain $r\La_{\rm grav} \sim 160$.
This is sufficient to suppress gravitational loop effects, and
also suppresses flavor-changing contributions from string/M-theory
states at the scale $\La_{\rm gauge}$.
We have also checked that the contact terms from bulk gauge fields
are negligible.
These estimates are quite rough, but we conclude that it is
very plausible that this mechanism can give a sufficiently large
radius so that anomaly mediation dominates.

\section{Conclusions}
We have studied a five-dimensional model with brane-localized visible
and hidden sectors localized on `3-branes' and shown that when the
compactification radius is properly stabilized, the transmission of
supersymmetry breaking to the visible sector proceeds by the mechanism of
anomaly-mediation.
Although the radius modulus participates strongly in
the supersymmetry breaking, it does not contribute to soft visible sector
masses at order $1/M_4^2$ because it does not directly couple to the
visible brane.
The stabilization mechanism for the radius modulus employed in this
paper is very simple, involving gaugino condensates in the bulk and on a
brane.
The bulk gauge fields do not give additional contributions to visible soft
masses due to the constraints of gauge invariance.
The advantage of this mechanism is that it gives a non-perturbative
superpotential for the modulus arising from field-theoretic mechanisms
that are under theoretical control.
It is also possible that that such a superpotential could also arise from
non-perturbative string/M-theory effects due to extended states.

This work is evidence that anomaly-mediated supersymmetry breaking
gives a model-independent contribution to soft supersymmetry breaking
in the visible sector at order $1/M_4^2$ in \emph{any} model with
SUSY breaking on a hidden-sector brane, and stabilized moduli.
If there are no additional light bulk fields that give a larger
contribution, anomaly-mediation dominates, giving a natural solution
to the supersymmetric flavor problem as well as potentially testable
predictions.
These features can be upset by the presence of additional bulk fields
with significant couplings to the visible sector.%
\footnote{For a phenomenologically interesting example, see \Ref{U1}.}
Knowledge of the true string theory vacuum, or experiment, is
required to find out if such light non-minimal bulk fields are
present.

\section*{Acknowledgments}
The authors would like to thank the Institute for Theoretical Physics at
Santa Barbara for hospitality during the course of this work.
We thank J. Bagger, T. Banks, J. Louis, M. Peskin, E. Poppitz,
L. Randall and F. Zwirner for comments and discussions on this work.
M.A.L. was supported by the National Science Foundation under
grant PHY-98-02551, and by the Alfred P. Sloan Foundation. R.S was supported
by the National Science Foundation under grant PHY-9870115. This research was
  also supported in part by the National Science Foundation under grant
PHY94-07194.

\end{document}